\documentclass{article}
\usepackage{graphicx}
\usepackage{amsmath}

\begin{document}

\title{MEDEA: a real time imaging pipeline for pixel lensing}
\author{Gerardo Iovane\thanks{%
E-mail geriov@sa.infn.it} \\
{\small \em Dip. di Ing. dell'Informazione e Mat. App., Universit\`a di Salerno,
Italy,}\\
{\small \em GNFM, Gruppo Nazionale di Fisica Matematica, Salerno,
Italy.}\\\\
Salvatore Capozziello\thanks{%
E-mail capozziello@sa.infn.it}\\
{\small \em Dipartimento di Fisica "E.R.Caianiello", Universit\`a di Salerno and}\\
{\small \em INFN Sez. di Napoli, Gruppo Collegato di Salerno,
Italy}.\\\\
Giuseppe Longo\thanks{%
E-mail longo@na.infn.it}\\
{\small \em Dipartimento di Scienze Fisiche, Universit\`a Federico II di Napoli, Italy, and}\\
{\small \em INAF Sez. di Napoli, Osservatorio Astronomico di Capodimonte, Napoli, Italy, and}\\
{\small \em INFN Sez. di Napoli, Universit\`a Federico II, Napoli, Italy}.\\
 }
\date{}

\maketitle

\begin{abstract}
Pixel lensing is a technique used to search for baryonic
components of dark matter (MACHOs) and allows to detect
microlensing events even when the target galaxies are not resolved
into individual stars. Potentially, it has the advantage to
provide higher statistics than other methods but, unfortunately,
traditional approaches to pixel lensing are very demanding in
terms of computing time. We present the new, user friendly, tool
MEDEA (Microlensing Experiment Data-Analysis Software for Events
with Amplification). The package can be used either in a fully
automatic or in a semi-automatic mode and can perform an on-line
identification of events by means of a two levels trigger and a
quasi-on-line data analysis. The package will find application in
the exploration of large databases as well as in the exploitation
of specifically tailored future surveys.
\end{abstract}

Pacs: 95.75.Mn, 95.75.Pq, 98.56.Ne, 98.62.Sb, 95.35.+d

\newpage

\section{Introduction}
During the last decade, much attention has been devoted to the
detection of MACHOs (Massive Astrophysical Compact Halo Objects).
Several teams have shown that microlensing can be successfully
used to detect the slight changes in the luminosity of background
stars caused by the passage of a massive deflector interposed
along (or close) the line of sight (Paczynski 1986, Alcock et al.
1993, Auborg et al. 1993, Udalski et al. 1993).

The pixel lensing method was proposed and implemented by the AGAPE
Collaboration to monitor simultaneously large numbers of stars and
therefore increase the probability to detect the intrinsically
rare microlensing events (Baillon et al. 1993, Ansari et al.
1997). A similar technique, based on image subtraction, was used
by the VATT-Columbia Collaboration (Tomaney and Crotts 1994,
Tomaney 1996). Both techniques were successfully tested; recently
the discovery of new candidate events towards the Andromeda galaxy
was announced (Auriere et al. 2001, Calchi Novati et al. 2002).
Additional microlensing candidates towards Andromeda were found by
Crotts (2000).

In pixel lensing, the search for microlensing events is performed
by monitoring the light curve of individual image pixels rather
than that of individual stars. Such approach, however, while
ensuring an enormous gain in the statistics, poses paramount
problems in resolving the lensed sources. From a computational
point of view, pixel lensing requires a large number of floating
point operations on each individual pixels. This leads to very
large computational loads and usually prevents the real-time
detection (and possible follow-up's) of the candidate events. In
this paper we describe the new tool MEDEA - already outlined in a
preliminary and much less complete version in (Capozziello and
Iovane 1999) - which, by making use of advanced data mining
techniques, allows the real time processing of pixel data.

This paper is organized as follows. In the Sect.2, we give the
scientific background of the  microlensing theory and of the pixel
lensing technique. Sect.3 is devoted to a general description of
the MEDEA environment. Sect.4 describes the Data Pre-Processing
and Sect.5 the Data Processing. The analysis of the data is
discussed in Sect.6, while some conclusions are drawn in Sect.7.

\section{Microlensing and  Pixel Lensing Technique}
If a massive object acting as a gravitational lens on a background
source is sufficiently close to the line of sight of the observer,
the light is deflected by an angle which is usually too small to
produce observable multiple images and we can observe only a
magnification of the flux coming from the source. The
magnification factor $A(t)$ is given by:
\begin{equation}
A(t)=\frac{u^{2}(t)+2}{u(t)\sqrt{u^{2}(t)+4}},
\end{equation}
where $u(t)=\theta _{s}D_{OL}/R_{E},$ $\theta _{s}$ is the angle
between the optical axis and the direction of the source, $D_{OL}$
is the distance
between the observer and the lens, ${\displaystyle R_{E}=D_{OL}\ast \theta _{E}=\sqrt{\frac{%
4GM}{c^{2}}\frac{D_{LS}D_{OL}}{D_{OS}}\text{ }}}$is the Einstein radius, $%
D_{LS}$ is the distance between the lens and the source; $D_{OS}$
is the distance between the observer and the source. These events,
also known as microlensing, are characterized by three main
features:
\begin{itemize}
\item[-]  \textbf{uniqueness of the event}: the probability that a
microlensing events occurs twice on the same star is assumed to be
zero; \item[-]  \textbf{symmetry of the light curve}: the light
curve (in point-like model of lens and source) has to be symmetric
about the maximum magnification point; \item[-]
\textbf{achromaticity of the event}: the ratio between the flux
variation in different colours, ${\displaystyle \frac{\Delta \phi
_{colour\_1}}{\Delta \phi _{colour\_2}}}$ is constant in time
(Schneider, Ehlers and Falco 1992).
\end{itemize}
It is useful to stress that a differential amplification of
extended sources can give rise to a chromatic, but still
symmetric, lensing curve (Han et al. 2000). To study the case of
pixel lensing, let us start from defining the flux inside a fixed
pixel as:

\begin{equation}
\phi ^{p}=\phi ^{\ast }+\phi _{bkg}+n,
\end{equation}

\noindent where $\phi ^{\ast }$ is the photon flux of the star
''at rest'' that will be lensed, $\phi _{bkg}$ is the photon flux
coming from neighboring stars falling in the same pixel, and $n$
is the noise. A microlensing event implies a variation of $\phi
^{p}$ with time:

\begin{equation}
\phi ^{p}=\phi _{bkg}+(A(t)-1)\cdot \phi ^{\ast }+n.
\end{equation}

\section{The Environment}
MEDEA is structured as follows (see Fig.1): (Iovane 2002)\footnote{%
Winner of the Price "Best Application of Measure and Automation in
Europe 2001-2002" (Assigned by National Instruments, Italy), 28
February 2002.}.

\noindent i) The Advanced Data Acquisition (A-DAQ) Unit:
responsible for the data acquisition and pre-reduction.

\noindent ii) The Control Unit (CU) piloting the Telescope Control
System (TCS) which drives the telescope following the instructions
provided either by the Data Base Control System (DBCS) or by the
user.

\noindent iii) The DataBase (DB) Unit: it is the ''intelligent''
part of the system where the data are stored  and processed
according to the simulations or previous observations (Jordan
1997).

\noindent iv) The Processing and Analyzing (P\&A) Unit: platform
where massive data analysis is performed.

\noindent v) Dispatcher Unit (DU):  which automatically builds a
status report on the different phases of the data flow. More in
detail: statistics, plots of data and events are produced and
stored by this module. Moreover, in the occurrence of special
events (such as an alert or failure of the system, or a short
microlensing event, for which a quick answer is needed) this unit
can reach and alert people in automatic mode thanks to an e-mail
service, SMS (Short Message System) and Fax Messages.

\noindent The Processing and Analyzing Unit consists of three main
units: the Data Pre-Processing Unit (DAPP) for astrometric
alignment and for photometric and seeing correction; the Data
Processing Unit (DAP) for peak detection of relevant luminosity
variation; the Data Analysis Unit (DAU) for best fits, color
correlations, $\chi ^{2}$ tests, Kolmogorov - Smirnov tests. The
A-DAQ Unit and DAPP Unit are organized in a fully automatic, non
interactive and on line modality. A First Trigger Level (for
selecting luminosity variations trough a peak detection algorithm)
is the most relevant component of DAP Unit. It consists of four
sections: a) peak detection procedure, b) star detection and
filtering algorithm, c) cosmic rays filter, and d) peak
classification (single, double, multiple peak curves) routines.
Also the DAP Unit operates in real time, thus implying that the
whole data reduction is performed while the next set of data is
acquired. The second trigger level, corresponding to the selection
of microlensing events, is inside the DAU. During this phase, we
also test whether the measured luminosity curves are compatible or
not with simple lensing models\footnote{Such models are single
point-like source and single point-like deflector model, double
point-like source and single point-like deflector model, extended
source with constant brightness and single point-like deflector
model.}

The events which pass the first trigger level and are incompatible
with the simple models included in the second trigger level are
studied off-line by means of interactive procedures. This study is
relevant in order to understand those events which are produced by
complex lensing phenomena (such as double deflector, planetary
system and so on), variable stars or novae and supernovae.

\section{Data Pre-Processing}

The Data Pre-Processing Unit (DAPP)\ is composed by three software
modules: Astrometric Alignment Unit, Photometric Correction Unit,
Seeing Correction Unit.

\subsection{Astrometric Alignment}

In the astrometric alignment module, we find three hierarchical
levels: the first one is responsible of the data I/O and controls
the user interface, the second one controls the learning of a
properly selected part of the reference image which will be
compared with the other images to be aligned, while the last
object performs the astrometric alignment according to the well
known transformation

\begin{equation}
\left(
\begin{array}{c}
x^{\prime } \\
y^{\prime }
\end{array}
\right) =\left(
\begin{array}{cc}
a_{11} & a_{12} \\
a_{21} & a_{22}
\end{array}
\right) \cdot \left(
\begin{array}{c}
x \\
y
\end{array}
\right) +\left(
\begin{array}{c}
\Delta x \\
\Delta y
\end{array}
\right)
\end{equation}

\noindent At the second level, it is possible to define a
dimension for the images on which the system has to learn the
pattern. In other words, the program fixes the size of the
calibration windows as a function of both the density of the field
and of its auto-similarity. Of course, it is possible to use the
full frame, but this could turn out to be computationally too
heavy. The structure of the learning phase is the following:

\begin{itemize}
\item  an automatic threshold filters the pixels having counts
above a fixed treshold $\phi_{th}$ (with
$\phi_{th}=\overset{-}{\phi }+3\sigma$);

\item  a clustering procedure builds pixel clusters and rejects
those clusters which are either too large or too small (compared
to a test performed on the pixel area), or have a very irregular
shape (evaluated against the inertial momenta of the pixel
cluster);

\item the evaluation of the center of mass for each surviving
cluster and of the corresponding pattern considered for the
astrometric calibration.
\end{itemize}

\noindent In order to perform pattern matching, we implemented a
tool (Image Advanced Interpreter Matching) which incorporates
image understanding techniques to interpret the template
information, and then uses this information to recognize the
template in the image. In order to generate information about the
features of a template image we have used the following
information source: a) geometric modeling of images; b) effective
non-uniform sampling of images; c) extraction of template
information.

This algorithm is built to fulfill the following main functions:

\qquad 1. Edge detection and clusters selection;

\qquad 2. Evaluation of geometrical parameters $x_{i}$: cluster's
area (in pixel), cluster's perimeter, number of holes in each
cluster, hole's area (in pixel), hole's perimeter, cluster's
inertial tensor (where the luminosity plays the role of the mass);

\qquad 3. Computation of a linear function of the previous
parameters for each cluster, accordingly to the formula:

\begin{equation*}
f^{Cluster}=\alpha _{1}x_{1}+\alpha _{2}x_{2}+...+\alpha _{n}x_{n}~,
\end{equation*}
where ${\displaystyle \alpha _{i}=\frac{1}{\sqrt{\left( x_{i}-\overline{x_{i}}\right) ^{2}%
}}}.$

\noindent The calibration is connected with the maximum
correlation between images, trough $f^{Cluster}$. In other words,
we minimize:

\begin{equation}
C=\sum_{j}(f_{j}^{i}-f_{j+1}^{l})^{2},
\end{equation}

\noindent where $j$ is the image index, while $i$ and $l$ are the
cluster indexes. We evaluate the coefficients $a_{ij}$ and $\Delta
x$, $\Delta y$ of eq.(4) by using a number of cluster larger than
the number of calibration parameters, in correspondence of the
minimum value of $C$. This approach reduces the amount of
information needed to characterize completely an image or pattern,
thus speeding up the searching process.

\subsection{Photometric Correction}
Changes in the observing conditions are locally corrected with
respect to the reference image which, by definition, we assume to
be the one with the best seeing. The algorithm, which was
implemented taking into account the possibility of parallel
computing, corrects local fluctuations of the flux due to
gradients in the image (e.g. the effect the lunar stray light).
The image is first divided in cells then, in addition to the pixel
coordinates $x_{i}$ and $y_{i}$, we have two cell indexes,
$x_{frame}$ and $y_{frame}$. Moreover, we have an index to select
the image, $I$. In this way, in order to select the luminosity of
a pixel in an image of the set we have to assign a vector with
five components:
$\left(I,~x_{frame},~y_{frame},~x_{i},~y_{i}\right) $ .

The mean flux value and the standard deviation are evaluated for
each cell. Noisy pixels, cosmic rays and bad pixels are rejected
and then the mean flux and the standard deviation are again
evaluated and stored. After these operations, we impose that the
mean value of the flux must be equal on the compared cells.

This method does not work properly for pixels lying on the edges of the cells\footnote{%
It is relevant to choose an optimized cell size in order to obtain
a good compromise between the field of view and the number of
pixels near the edges of the cells. For example, given an image
taken on Andromeda at a distance of 690 Kpc with a CCD camera with
$1024\times 1024$ pixels, a good cell size is one order smaller
than the original image size (i.e. about $100\times 100$).}. If at
the end of the analysis there is a pixel on the edge of a cell
with an interesting light curve, a specific photometric alignment
is performed off-line by posing the relevant pixel at the center
of a new cell and then iterating the above described procedure.

In the first prototype of the pipeline, we implemented a
bi-dimensional interpolation on edges, but the signal turned out
to be diluted by the interpolation process thus imposing the
introduction of the cell approach. Fig. 2, 3, and 4 show the
reference image, the current image, and the current image after
the photometric correction.

%
%

Moreover, we find the grey level histograms before and after
photometric calibration in Fig.5 and 6.

%

\subsection{Seeing Correction}
The real time (in a post processing point of view) correction of
the seeing is a relevant issue (cf. Sedmak 1999). The analysis is
performed on a cluster of pixels (superpixel) with a size large
enough to cover the PSF. In the standard pixel lensing data
reduction (Ansari 1997, Le Du 2000), the superpixel size is chosen
to be large enough to cover the worse PSF and is the same for all
images. Here we dynamically select the dimension of the superpixel
with respect to the PSF measured on some calibrators in the fields
(i.e. for each image corresponding to a given observation). The
seeing variation factor is evaluated and defined as the ratio of
the area in pixels of selected clusters between a reference image
and another one.

This number becomes the input to build the kernel $k_{ij}$ of a
morphing algorithm that in our case is a dilation algorithm. Then
if $K$ is the kernel and $P$ is the pixels matrix, the seeing
corrected image $\widetilde{P}$ is:

\begin{equation}
\widetilde{P}=\frac{\underset{a}{\sum }\underset{b}{\sum }\left(
K_{b}^{a}\otimes P_{a}^{b}\right) }{\underset{a}{\sum }\underset{b}{\sum }%
K_{b}^{a}}\qquad \text{with\qquad }
\begin{array}{c}
a\in \left[ i-l,i+l\right] \\
b\in \left[ j-m,j+m\right]
\end{array}
,
\end{equation}

where $l\times m$ is the kernel size for the convolution. In this
procedure, the kernel is a polynomial function in $\Re ^{2}$ and
the superpixel can be chosen between rectangular and exagonal one
(see Fig.7). In Figs. 8 and 9 we show the same field in different
seeing conditions.

%
%

\section{Data Processing}
The Data Processing Unit is made of four sub-units: Star Detection
and filtering Unit, Cosmic Detection and filtering Unit, Peak
Detection Unit, and Peak Classification Unit. The cosmic rays unit
just switches off the saturated pixels, while the peak
classification unit splits the curves in function of the number of
the peaks attributing them to different classes.

\subsection{Star Detection}
This component finds and rejects resolved objects inside the
field. After performing several tests, we decided to work in the
transformed space. In astronomical image processing the Fourier
Transform is often used to select the main characteristics or the
morphology of bright objects (Pratt 1977, Sedmak et al. 1981).
If we consider the image as a function of two variables $%
f(x,y)\in C^{2}$, \ we will be able to define a transformed image
$\widehat{f}(u,v)\in C^{2}$. In our case $\widehat{f}(u,v)$ is
obtained by fast Fourier transforming the image $f(x,y)$

\begin{equation}
\widehat{f}(u,v)=\frac{1}{NM}\overset{N-1}{\underset{x=0}{\sum }}\overset{M-1%
}{\underset{y=0}{\sum }}f(x,y)\cdot e^{-2\pi i\left( \frac{ux}{N}+\frac{vy}{M%
}\right) },
\end{equation}

\noindent where $(u,v)$ are the horizontal and vertical spatial
frequencies. The function $\widehat{f}(u,v)$ is a complex image in
which the high frequencies are clustered at the center, while the
low ones are located at the edges. Fig.10 shows the complex image
$\widehat{f}(u,v)$ corresponding to the real image $f(x,y)$ shown
in Fig.2.


To select and reject the resolved objects - which correspond to
low frequency signals in the transformed image - we implemented an
adaptive high pass filter in the transformed space which produces
an automatic thresholding in the frequencies domain. In Figs. 11
and 12, we give the scheme of the filter and its effect on the
complex image, while in Fig.13 we show the same image
antitransformed and therefore cleaned of the resolved objects.

%


\subsection{Peak Detection}
The detection of the luminosity variation in the light curve is
performed by a peak detection algorithm. First of all we evaluate
the background level as:

\begin{equation}
\phi _{bkg}=\min \left( \frac{1}{\alpha +1}\overset{j+\alpha }{\underset{i=j%
}{\sum }}\phi _{i}\right) \qquad \forall j=1,...,n-\alpha +1,
\end{equation}

\noindent where $n$ is the number of points in the light curve
(e.g. the number of images), $\alpha +1$ is the dimension of the
window which we use to evaluate $\phi _{bkg}$ ($\alpha +1<n$). We
evaluate the mean value of the flux in a window running on the
light curve and having a size of $\alpha +1$; then we consider as
background the minimum of the means and then evaluate the maximum
standard deviation $\sigma $ to assess its stability. Before the
peak detection step, we apply  a median filter on the signal in
order to estimate the best peak parameter by separating pure
signals from noise fluctuations. If $Y$ represents the output
sequence, id est the filtered data, and if $J_{i}$ represents a
subset of the input sequence $X$ centered on the $\ i^{th}$\ element of $x$
\begin{equation*}
J_{i}=\left\{
x_{i-r},\,x_{i-r+1},\,...,\,x_{i-1},\,x_{i},\,x_{i+1},\,...,\,x_{i+r-1},%
\,x_{i+r}\right\} ,
\end{equation*}

\noindent if the indexed elements outside the range of $X$ are
equal to zero, the function gives the elements of $Y$ by using:

\begin{equation*}
y_{i}=Median\left( J_{i}\right) \qquad \forall i=0,\,1,\,2,...,\,n-1,
\end{equation*}

\noindent where $n$ is the number of elements in the input
sequence $X$, and $r$ is the filter rank. The effects of the
filter are shown in Fig.14.


\noindent Afterwards we perform a peak detection on the filtered
signal with a polynomial (usually of the 2$^{-nd}$ order is
enough) linearly combined, trough a sum operator:
%
%

\begin{equation}
\underset{i\in I}{\biguplus }\alpha _{i}x^{2}+\beta _{i}x+\gamma _{i},
\end{equation}

\noindent where $I$ \ is the space of peaks and the symbol
$\biguplus $ means a sum of functions at different $x$. For
instance, for a 2-modal light curve, we have:

\begin{equation}
\underset{i\in I}{\biguplus }\alpha _{i}x^{2}+\beta _{i}x+\gamma
_{i}=\left\{
\begin{array}{c}
\alpha _{1}x^{2}+\beta _{1}x+\gamma _{1}\qquad \forall x\leq x^{\ast } \\
\alpha _{2}x^{2}+\beta _{2}x+\gamma _{2}\qquad \forall x\geq x^{\ast }
\end{array}
\right. .
\end{equation}

\noindent Outputs of this routine are the parameters to be used
for the fit of the light curve: i.e. the number of peaks, the
location ($t_{0}$) and the amplitude.

\section{Data Analysis}
The on-line and off-line are performed in the Data Analysis Unit.
Here, the statistical $\chi ^{2} $ and Kolmogorov-Smirnov tests,
and the evaluation of a specific quality factor are made. The
color correlation among light curves is tested in different color
bands at the end of the process.

\subsection{On-line and Off-line Analysis}

The on-line sub-unit\footnote{This unit was tested on data
collected on the 1.3 meter McGraw-Hill Telescope, at the MDM
observatory, Kitt Peak (USA) and actually it is in a fine tuning
phase for data to be collected at the TT1 (Toppo Telescope)
CastelGrande (Italy).} can perform basic fits using the parameters
derived in the peak detection phase by means of the
Levenberg-Marquardt algorithm for non-linear fit. This sub-unit is
fully automated and perform in real time the following fits:

\begin{itemize}
\item  \textbf{{Paczynski test.}} We consider magnification $A$ defined as

\begin{equation}
A(t)=\frac{u^{2}(t)+2}{u(t)\sqrt{u^{2}(t)+4}}\sim \frac{1}{u(t)}=\frac{1}{%
\sqrt{u_{0}^{2}+\left( \frac{t-t_{0}}{t_{E}}\right) ^{2}}},
\end{equation}

\noindent with $u_{0}=r_{0}/R_{E}$ , where $r_{0}$ is the impact
parameter at maximum
magnification, $t_{0\text{ }}$is the time of maximum magnification, and%
\textbf{\ } $t_{E}$ is the Einstein time.


%
\item  \textbf{{Double source test}. } According to the model by
Griest and Hu (1992) there will be two $t_{0}$ and two $u_{0}$ and
therefore two functions: $u_{1}(t_{i})$\ and $u_{2}(t_{i})$.
Consequently, let $\phi _{1}$ and $\phi _{2}$ be the flux of the
two parts and the flux offset ratio $\omega =\frac{\phi
_{2}}{\phi_{1}+\phi _{2}}$, then the light magnification is given
by:

\begin{equation*}
A_{bs}(u_{1}(t),\,u_{2}(t))=(1-\omega )A(u_{1}(t))+\omega A(u_{2}(t)).
\end{equation*}

\noindent The light curve is the superposition of two light curves
for point-like sources, affected by a point-like mass deflector.

\item  \textbf{{Extended circular source}. } This is the case of
an optical system with a point-like lens and an extended circular
source having radius $r$ and constant surface brightness. The
implementation uses the model proposed by Witt and Mao (Witt and
Mao 1994).
\end{itemize}

\noindent MEDEA also contains a set of tools to study more complex
events (cf. Di Stefano and Mao 1996, Dominik 1998, Dominik 1999).
In particular, the events which pass the first trigger level but
are not selected by the second trigger can be analyzed off-line
both automatically and manually. The analysis is performed by
means of the comparison with reference simulated data stored
inside the database. The DB events are simulated in order to test
the following hypotheses: a) double point-like lenses, and single
point-like source; b) nova and supernova\footnote{For novae and
supernovae events, a MEDEA subunit was tested on a light curve
analyzed by SLOTT-AGAPE collaboration (Calchi Novati et al. 2002).
The light curve of these events are characterized by a very strong
and chromatic flux variation, that has been considered as due to
novae (Modjaz and Li 1999, Riffeser et al. 2001).}; variable
source.

Tools for a traditional analysis (i.e. standard light curves
analysis) are implemented too\footnote{More details on the
off-line analysis and other MEDEA modules will be given in a
companion and successive paper.}.

\subsection{Selection of lensing events}
The second trigger consists of two selective procedures: the first
one is a statistical phase, while the second one is specific for
microlensing. The $\chi ^{2}$ and Kolmogorov-Smirnov test are
implemented to evaluate the quality of the tested hypotheses. For
each model we consider the $\chi ^{2}$ test and the $Q$-factor of
Kolmogorov-Smirnov test, then a quality factor M is estimated. It
is defined as

\begin{equation}
M=Q\times \chi ^{2},
\end{equation}

\noindent and light curves with $M\leq 1.5$ are considered as
possible microlensing
events\footnote{%
Following the analysis and the results in (Calchi Novati et al.
2002), in order to obtain a better confidence on the sample, we
also implemented a code to perform the Durbin Watson statistical
test (Durbin and Watson 1951).}. The events, passing the previous
tests, are considered for the color test. The cross correlation
between different colors is estimated as:

\begin{equation}
C=\frac{\overset{N}{\underset{n=1}{\sum }}(\phi _{n}^{p}-\left\langle \phi
^{p}\right\rangle )_{color\_1}\times (\phi _{n}^{p}-\left\langle \phi
^{p}\right\rangle )_{color\_2}}{\sqrt{\overset{N}{\underset{n=1}{\sum }}%
(\phi _{n}^{p}-\left\langle \phi ^{p}\right\rangle )_{color\_1}^{2}\times
(\phi _{n}^{p}-\left\langle \phi ^{p}\right\rangle )_{color\_2}^{2}}},
\end{equation}

\noindent where the index $p$ is linked with pixel and $n$ with
the point along the light curve. For $C\geq 0.98$ the event is
considered a good microlensing candidate.

\section{Summary and conclusions}

MEDEA was specifically tailored to perform automatic microlensing
search. Most of the tools presented in this paper can be applied
in other fields, where pipelines and data mining procedures are
needed for large amounts of data. The procedures implemented in
MEDEA environment and presented in the Data Acquisition (DAQ), and
in the Data Pre-Processing (DAPP) units have been tested on
simulated data and images, while the Data Processing (DAP) and the
Data Analysis (DAU) units have been tested on a set of data
collected at the 1.3 m McGraw-Hill telescope, at MDM observatory -
Kitt Peak, in the period from the end of September to the end of
December 1999, by using the bulge of Andromeda as target. The
results agree with ``traditional'' pixel lensing analysis
performed by the AGAPE Collaboration. The flexibility of the
system allows to perform automatic, semiautomatic or researcher
assisted procedures of the candidate microlensing events. Common
microlensing events due to single and double point-like source, or
extended source are analyzed on-line and with automatic
procedures, while more complex events such as double lens, novae,
supernovae, variable stars or multiple deflectors system
(planetary systems), double point-like sources and lenses are
studied off-line, either with automatic procedures (based on the
use of a database of simulated events), or individually with other
tools. In this way, all the advantages of non-automatic procedure
are kept, and also common events are automatically studied so that
the researcher has only to control them. The detection of short
events or events with huge main peaks and secondary ones near the
first (as in the case of planetary system) becomes possible thanks
to the real time light curve monitoring and to the dispatcher
implementation. In addition, thanks to the fast- and real-time
analysis, the exact knowledge of the time region in which there is
a luminosity magnification gives the possibility to use larger
telescopes to resolve the pixels and obtain more detailed
astrophysical information on the magnified star. Finally the
reduction of time to spend for data analysis after data taking is
a good starting point to perform lensing analysis on large field
CCD data obtained in the course of specific surveys.
\\\\
Acknowledgements: The authors wish to thank SLOTT-AGAPE Collaboration for
useful suggestions and comments about lensing, and G.Sedmak for suggestions
about the seeing correction. This work was partly sponsored by the Ministero
Italiano per l'Universit\'a e la Ricerca scientifica e tecnologica in the
framework of a COFIN 2000.

\newpage

\newpage

\begin{center}
{\Large Legends to Figures}
\end{center}

Figure 1. The MEDEA flow chart. The dotted part (e.g. Telescope
Remote Control, Telescope Remote Observing) in the control unit
could be very useful if one starts to think about remote control
and observations.\\

Figure 2. ANDROMEDA in a good photometric condition.\\

Figure 3. ANDROMEDA in a not so good photometric condition.\\

Figure 4. ANDROMEDA prototype image after photometric alignment.\\

Figure 5. Gray level histogram of two images before photometric
alignment.\\

Figure 6. Gray level histogram of two images after photometric
alignment.\\

Figure 7. Rectangular and exagonal superpixel frame.\\

Figure 8. Image with a good seeing condition (VLT Image of Distant
Galaxies in AXAF Deep Field by ESO 2000).\\

Figure 9. Image with a worst seeing condition (VLT Image of
Distant Galaxies in AXAF Deep Field by ESO 2000).\\

Figure 10. FFT transformed image of ANDROMEDA.\\

Figure 11. The scheme of the filter in the transformed space.\\

Figure 12. The filtering effect on ANDROMEDA image in transformed
space.\\

Figure 13. The filtering effect on ANDROMEDA image in
anti-transformed space (or real).\\

Figure 14. Effect of median filter.

\end{document}